# Chiral Metafilms and Surface Enhanced Raman Scattering For Enantiomeric Discrimination of Helicoid Nanoparticles.

*Martin Kartau[a], Anastasia Skvortsova[b], Victor Tabouillot[a], Rahula Kumar[a], Polina Bainova[b], Vasilii Burtsev[b], Vaclav Svorcik[b], Nikolaj Gadegaard[c], Sang Won Im[d], Marie Urbanova[b], Oleksiy Lyutakov[b*], Malcolm Kadodwala[a*], Affar S. Karimullah[a*].*

[a] *School of Chemistry, Joseph Black Building, University of Glasgow, Glasgow, G12 8QQ, UK*

[b] *Department of Solid State Engineering, University of Chemistry and Technology, 16628 Prague, Czech Republic*

[c] *School of Engineering, Rankine Building, University of Glasgow, Glasgow, G12 8QQ, UK*

[d] *Department of Materials Science and Engineering, Seoul National University, Seoul, 08826, Republic of Korea*

______________________________

## Abstract

Chiral nanophotonic platforms provide a means of creating near fields with both enhanced asymmetric properties and intensities. They can be exploited for optical measurements that allow enantiomeric discrimination at detection levels > 6 orders of magnitude than is achieved with conventional chirally sensitive spectroscopic methods based on circularly polarized light. The optimal approach for exploiting nanophotonic platforms for chiral detection would be to use spectroscopic methods that provide a local probe of changes in the near field environment induced by the presence of chiral species. Here we show that surface enhanced Raman spectroscopy (SERS) is such a local probe of the near field environment. We have used it to achieve enantiomeric discrimination of chiral helicoid nanoparticles deposited on left and right-handed enantiomorphs of a chiral metafilm. "Hotter" electromagnetic hotspots are created for matched combinations of helicoid and metafilms (left-left and right-right), while mismatched combinations leads to significantly "cooler" electromagnetic hotspots. This large enantiomeric dependency on hotspot intensity is readily detected using SERS with the aid of an achiral Raman reporter molecule. In effect we have used SERS to distinguish between the different

EM environments of the plasmonic diastereomers produced by mixing chiral nanoparticles and metafilms. The work demonstrates that by combining chiral nanophotonic platforms with established SERS strategies new avenues in ultrasensitive chiral detection can be opened.

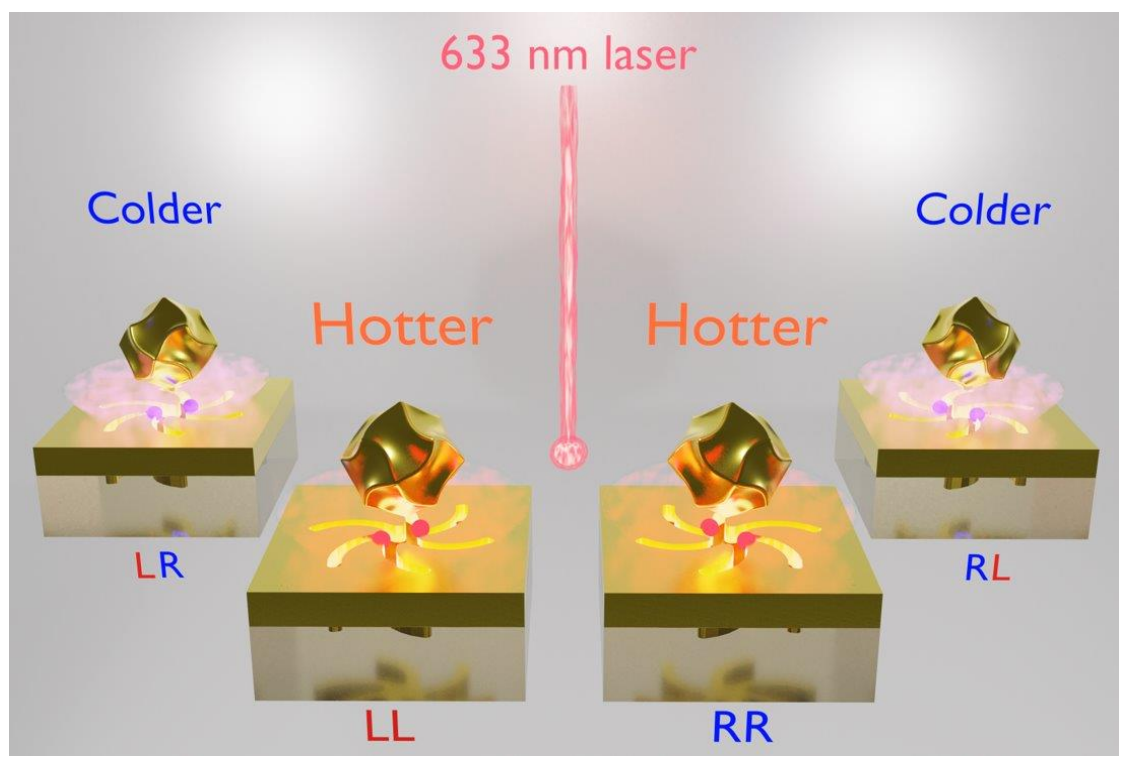

Table of Contents Graphic

## Introduction

Chiral organic and inorganic materials have broad impact in science and technology spanning research areas as diverse as pharmaceutical to quantum computing. Chirally sensitive, chiroptical spectroscopic methods based on the differential interaction of circular polarized light, such as electronic circular dichroism (CD), are important tools since they provide rapid post synthesis characterization of materials.[1] However, the inherent weakness of CD results in it being far less sensitive than non-chirally sensitive measurements.[2] Nanophotonic platforms such as plasmonic / dielectric metamaterials provide strong light-matter interactions that have successfully been utilized to increase the sensitivity in biosensing [3-5] and Surface Enhanced Raman Optical Activity (ROA) SEROA.[6-8] Nanophotonic platforms provide novel strategies for enhancing the sensitivity of chiral detection to (sub) monolayer quantities of the materials (≤ femtomole).[6-10] Nanophotonic platforms provide novel strategies for enhancing the sensitivity of chiral detection to (sub) monolayer quantities of the materials (≤ femtomole).[9-13] There are two different philosophies for exploiting nanophotonic platforms which will for brevity be termed the "direct" and "indirect" strategies. The direct strategy can be considered an extension of conventional CD methods, achieveophotonic platforms are used to amplify both the intensities and chiral asymmetries of electromagnetic (EM) fields, to enhance dichroic responses.[10, 14] This is an elegant and simple concept; however, it does have weaknesses which may limit potential applications, namely, the near UV region <230 nm) that is important in CD for biological and chemical samples as it is challenging to achieve with nanophotonic platforms that are compatible with aqueous environments. The indirect strategy is a complete divergence from conventional chiroptical spectroscopic paradigms. It relies on detecting asymmetric changes in the response of left and right-handed (LH and RH) chiral plasmonic structures on the introduction of a chiral medium.[15] This differential response is due to chiral media inducing asymmetric changes in the intensities and chiral properties of the near fields of LH and RH substrates.[15] Since the indirect strategies do not rely on the nanophotonic structures being in resonance with an electronic transition of a molecule, gold plasmonic structures can be used, which widen the range of systems, both biological and chemical, which can be studied.[15] However, this approach does also have inherent weakness, primarily the complexity of the phenomena responsible for the observed effects. For instance, chiral detection is strongly dependent on the orientational order of the adsorbed target molecule, and it is, surprisingly, dependent on the charged state of the molecule, indicating a capacitive contribution of the response which is not accounted for previous EM numerical modelling. [15, 16] An issue with some early studies employing the indirect strategy, was that local changes in near field properties were monitored by observing small changes in a far field optical response. Recently, it has been shown that directly probing changes in the local EM environment caused by chiral media with luminescence measurements, larger asymmetric responses can be achieved than with far field measurement methods. [17, 18]

Here we demonstrate how surface enhanced Raman scattering (SERS) can be used to detect large asymmetric local changes in near field intensities of enantiomorphic metafilms, containing chiral nanocavities, induced by the deposition of chiral helicoid nanoparticles. Thus, providing a novel approach for the enantiomeric discrimination of plasmonic nanoparticles. The origin of the effect is the differential intensities of hot spots generated in gap regions formed where the helicoid and chiral metafilms are very close. Together, an achiral molecule, biphenyl-4,4'-dithiol (BPDT), acts both to facilitate binding between the helicoid nanoparticles and the metafilm, and as a Raman reporter of the near field intensities. This study provides a framework of using an alternative indirect approach, based on SERS, for exploiting nanoplatforms for chiral sensing.

## Results and Discussion

In this study we combine Au metafilms with chiral nanoparticles. The Au metafilms have multiple nanostructured arrays consisting of six armed "shuriken" shaped nano-indentation that are intrinsically either left-handed (LH) or right-handed (RH), Figure 1 (A) (Details in Supporting Information Figure S1). The substrates are based on nanopatterned polycarbonate templates, which have been described in detail elsewhere.[16, 19, 20] The metafilms are then formed by coating the templates with a 100 nm Au film and have previously been used in chirally sensitive biosensing to study structural changes in proteins, high-throughput nanoscale spatial functionalization strategies, and active plasmonics for controllable chiral optical response. [4, 21] Recently, they have also been used to demonstrate how near field effects can enhance probing chiral media with higher sensitivity using functionalized quantum dots and measuring the plasmonically enhanced luminescence.[17] Reflectance spectra collected from shurikens in air, display multiple resonances over the visible and near infrared spectral region, Figure 1 (D), with sharper features between 700-770 nm region displaying Fano coupling between multiple modes, and a broader resonance between 500-650 nm.[16, 17] This latter feature is associated with a resonance that has a bisignate line shape in optical rotation dispersion (ORD). As would be expected due to the chiral nature of the shurikens, LH and RH structures show equal and opposite ORD spectra.

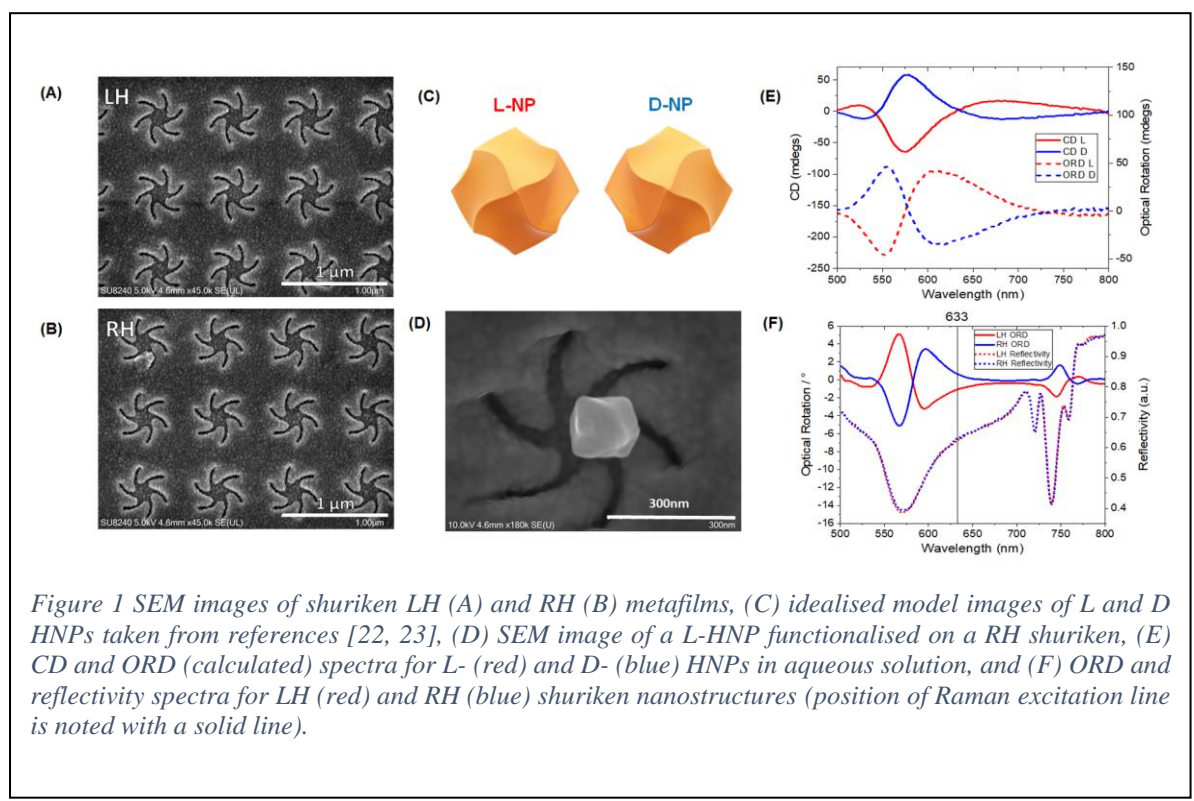

*Figure 1 SEM images of shuriken LH (A) and RH (B) metafilms, (C) idealised model images of L and D HNPs taken from references [22, 23], (D) SEM image of a L-HNP functionalised on a RH shuriken, (E) CD and ORD (calculated) spectra for L- (red) and D- (blue) HNPs in aqueous solution, and (F) ORD and reflectivity spectra for LH (red) and RH (blue) shuriken nanostructures (position of Raman excitation line is noted with a solid line).*

Rhombic dodecahedral shaped, chiral nanoparticles, subsequently referred to as helicoidal nanoparticles (HNP), were made using an established synthesis strategy, Figure 1 (B).[22, 23] Seed-mediated colloidal growth using an amino acid as the shape modifier helps control the handedness and chiral plasmonic resonance of the nanoparticles. In this work, chirality was conveyed on to the HNP by the presence of a chiral ligand, cysteine (L- or D-), with the absolute configuration of the particle being controlled by that of the ligand (Supporting Information Figure S2). Consequently, particles generated with L-cysteine as an additive are termed L-HNPs and those generated using D – cysteine are termed D – HNPs. The circular dichroism (CD) response from these plasmonic nanoparticles is highest in the 500-650 nm spectral region, Figure 1 (E), overlapping with that shown by the shuriken metafilms. Functionalization of the HNPs to the metafilm is performed by first generating a self-assembled monolayer (SAM) using BPDT on the Au films. The adsorption of BPDT caused red shifts of ~ 2.2 nm on both LH and RH shuriken metafilms, indicative of no significant difference in coverage between the enantiomorphs. The nanoparticles are then functionalized using the second thiol group of the now bound BPDT and any weakly bound HNPs are subsequently rinsed away. The combination of HNP and shuriken can be regarded as a diastereomer. As a convention, a two-lettered labeling system is used to describe the different diastereomeric combinations of shuriken and HNP, 'LL', 'LR', 'RL', and 'RR', where the first letter describes the handedness of the shuriken nanostructure and the second letter the handedness of the nanoparticle. For simplicity D-HNP is considered a right-handed particle and referred to by 'R'. The HNP deposition process produced structurally heterogeneous layers with particles being immobilized at random positions both within the shuriken indentation and in the regions between structures, Figure 2 (A) (Supporting Information Figure S4 for higher magnification micrographs). The deposition of HNPs induced broadening in the reflectivity, Supporting Information Figure S3, indicating coupling between the HNPs and metafilms,[24] and red shift the bisignate ORDs, Figure 2 (B), by ~1.6 nm for all diastereomeric combinations, indicative of both an absence of asymmetry in the far field linear response and similar coverages of HNP for the four combinations. SERS response of the achiral BPDT was measured and to mitigate any possible variations in HNP coverage on metafilms, the Raman measurements were performed on multiple sites of each of the L- or D- HNPs coated substrates. Each substrate has 9 pairs of LH and RH shuriken arrays, Supporting Information Figure S5, and Raman measurements were performed on all pairs to collect 9 spectra of each diastereomeric combinations (36 in total), Figure 3 (A)-(D). The solid lines show the means of the Raman measurements, and the standard errors are shown by the shaded bands surrounding each solid line. Similar spectra, in terms of positions of bands and relative intensities, were obtained from all substrates, and they are solely dominated by bands associated with BPDT. The BPDT SERS spectra in this work are very similar to those obtained previously from BPDT immobilized on either Au nanoparticles or commercial SERS (Klarite) substrates.[25] They display characteristic strong Raman peaks at ~1589, 1285, 1200, and 1084 cm$^{-1}$, vibrational assignments are given in Supporting Information Table S1.

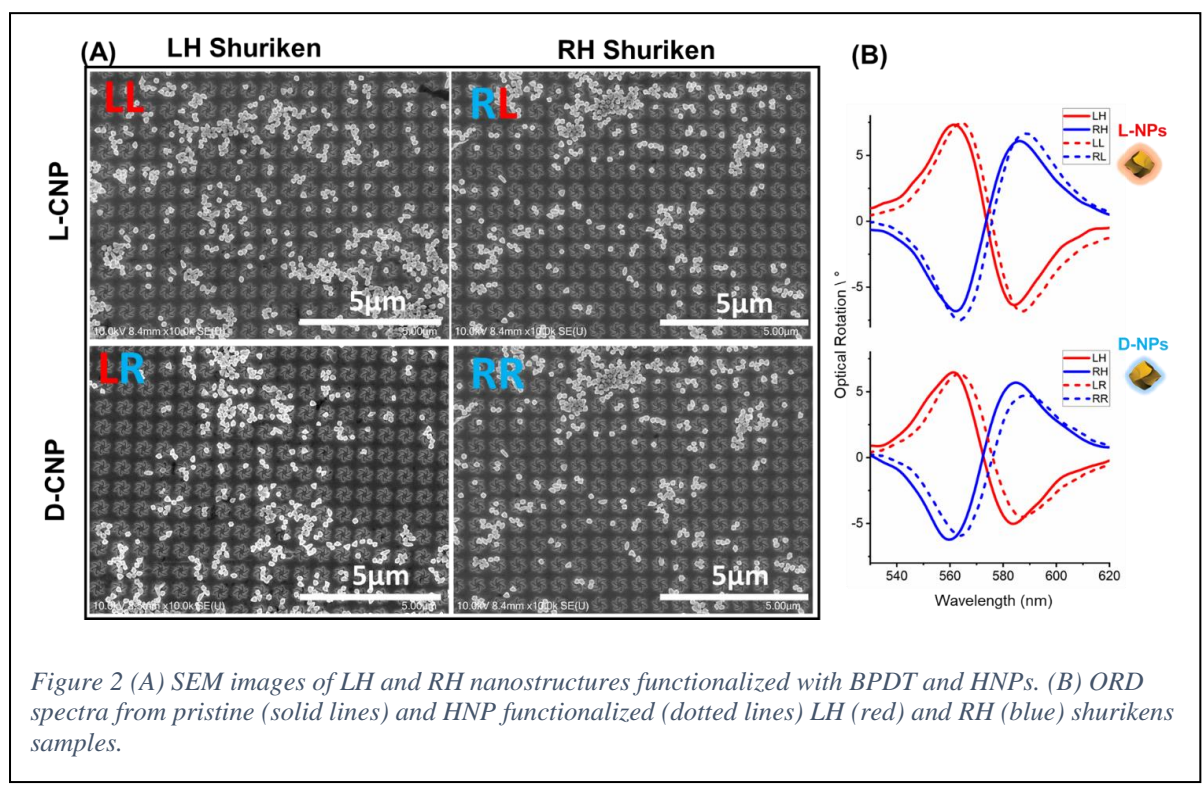

*Figure 2 (A) SEM images of LH and RH nanostructures functionalized with BPDT and HNPs. (B) ORD spectra from pristine (solid lines) and HNP functionalized (dotted lines) LH (red) and RH (blue) shurikens samples.*

No Raman spectra were detectable from BPDT layers on unstructured Au films in the absence of HNPs, Figure 3 (A). As references, Raman spectra were collected from BPDT functionalized LH and RH shurikens in the absence of HNPs, Figure 3 (A), and from L- and D-HNPs immobilized on BPDT layers on unstructured Au films, Figure 3 (B). L- and D-HNPs on unstructured Au surfaces give Raman spectra which, within experimental error, have near identical intensities. The Raman spectra obtained from BPDT from the two shuriken enantiomorphs, that is shurikens and HNPs, Figure 3 (C)-(D), have comparable intensities, where the slight differences between the two samples can be attributed to the large level of heterogeneity of the deposited HNP layers.

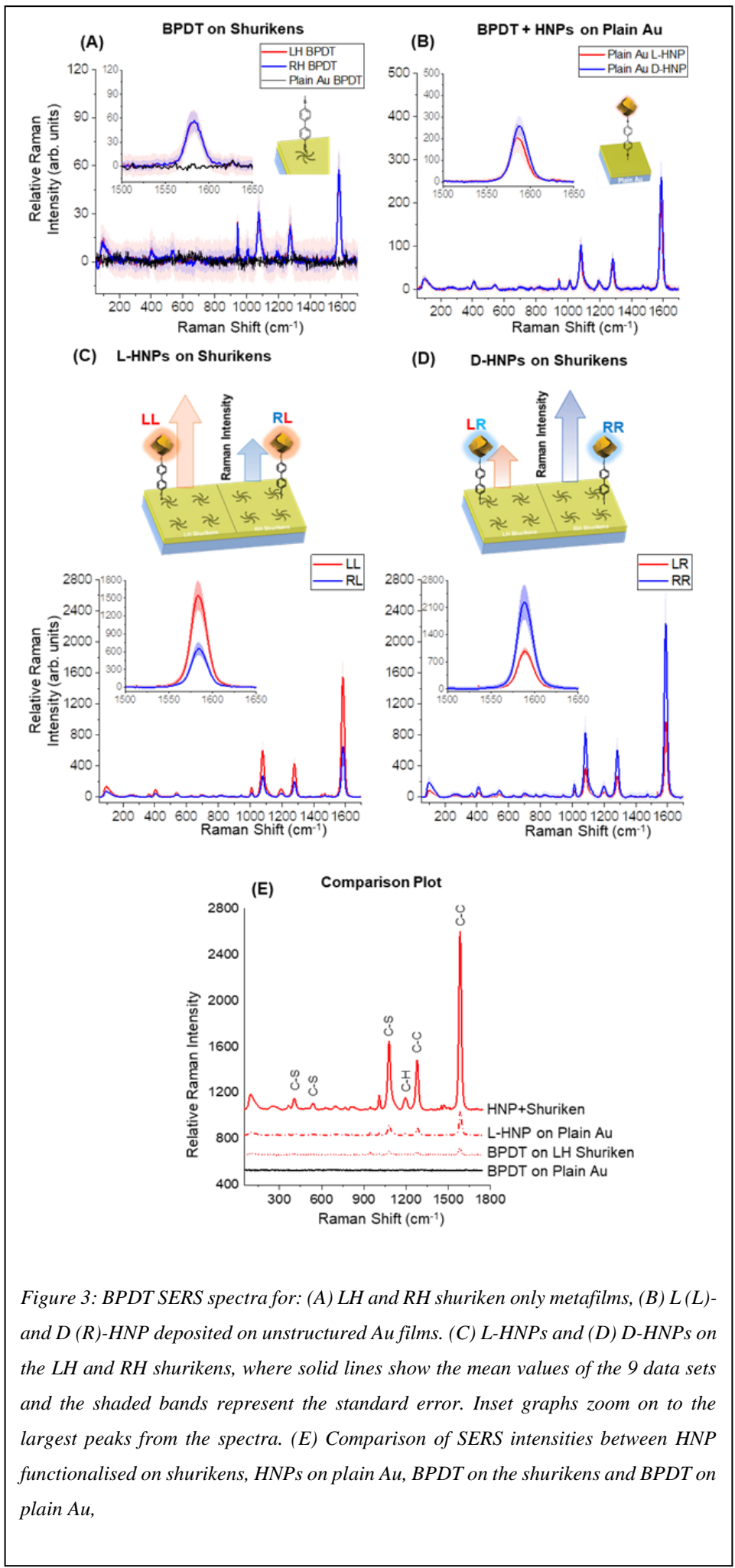

*Figure 3: BPDT SERS spectra for: (A) LH and RH shuriken only metafilms, (B) L (L)- and D (R)-HNP deposited on unstructured Au films. (C) L-HNPs and (D) D-HNPs on the LH and RH shurikens, where solid lines show the mean values of the 9 data sets and the shaded bands represent the standard error. Inset graphs zoom on to the largest peaks from the spectra. (E) Comparison of SERS intensities between HNP functionalised on shurikens, HNPs on plain Au, BPDT on the shurikens and BPDT on plain Au,*

The central result of this study is the observation of the dependencies of the overall intensities of the Raman spectra on the symmetry properties of the diastereomer combinations of HNP and shuriken. In Figure 3 (C) a comparison of spectra from L-HNP on both shuriken enantiomorphs is shown, while the equivalent comparison for the D-HNP (R) is shown in Figure 3 (D). The SERS spectra clearly show that the matched combinations, LL and RR, of nanostructures and HNPs result in higher Raman signal intensities compared to their mismatched counterparts, LR and RL. The average ratios of the intensities of bands (derived from

~1084, 1285 and 1589 cm$^{-1}$ bands) for LR / RR and RL / LL combinations are 0.43 and 0.46 respectively.

A final point about the Raman data is the dependency of the magnitudes of the responses on the type of substrates, Figure 3 (E). The following trend in the intensities of the BPDT Raman signal is observed; HNP + Shurikens (Average) > HNP + Flat substrate > Shuriken only. Specifically, average signal from the HNP + Shuriken is ~ 7.5 time greater than that from HNP on unstructured Au; which is in turn ~ 4.5 times greater than the response from BPDT on shurikens. The observation that shuriken only structure gives the weakest Raman response can be readily reconciled when one considers that these structures produce weaker hot spots than the other two cases. This is because the hotspots associated with the shuriken structures (without HNPS) are predominately in the arms. The width of the arms is much larger than the gap regions formed between the HNP and either the Au unstructured or shuriken surface.

We propose that the dependency of the Raman reporter signal on the symmetry properties of the diastereomeric combinations of HNP and shuriken arises from the different EM environments that they possess. It is well established that light-matter coupling is strongly enhanced in resonant cavities and antennae.[26, 27] Enhancement of spontaneous emission for instance can be engineered using Forster resonance energy transfer or the Purcell effect by placing the emitter in a resonant cavity generating large local density of optical states[28, 29]. In the case of enhanced Raman scattering from plasmonic cavities / antennae, it can be shown that the total observable enhancement is a cumulative result of field enhancement of the incoming radiation and the enhanced radiative decay rate at the Stokes frequency.[30, 31] The relative size of a SERS response is generally discussed in terms of an enhancement factor (EF), which is the relative size of a signal compared to that obtained from conventional Raman measurement. If the frequency of the Raman scattered light is close to that of the incident light, then,[32]

$$EF \approx \left|\frac{E_{loc}}{E_i}\right|^4$$

(1)

where $E_{loc}$ is the electric local field amplitude at the Raman active site (or hot spot) and $E_i$ is the amplitude of the incident excitation. Hence, Equation 1 relates the EF to intensities of hotspot regions in the EM environment. Consequently, our hypothesis is that,[7]

$$\left|\frac{E_{loc}}{E_i}\right|_{LL,RR} > \left|\frac{E_{loc}}{E_i}\right|_{LR,RL}$$

(2)

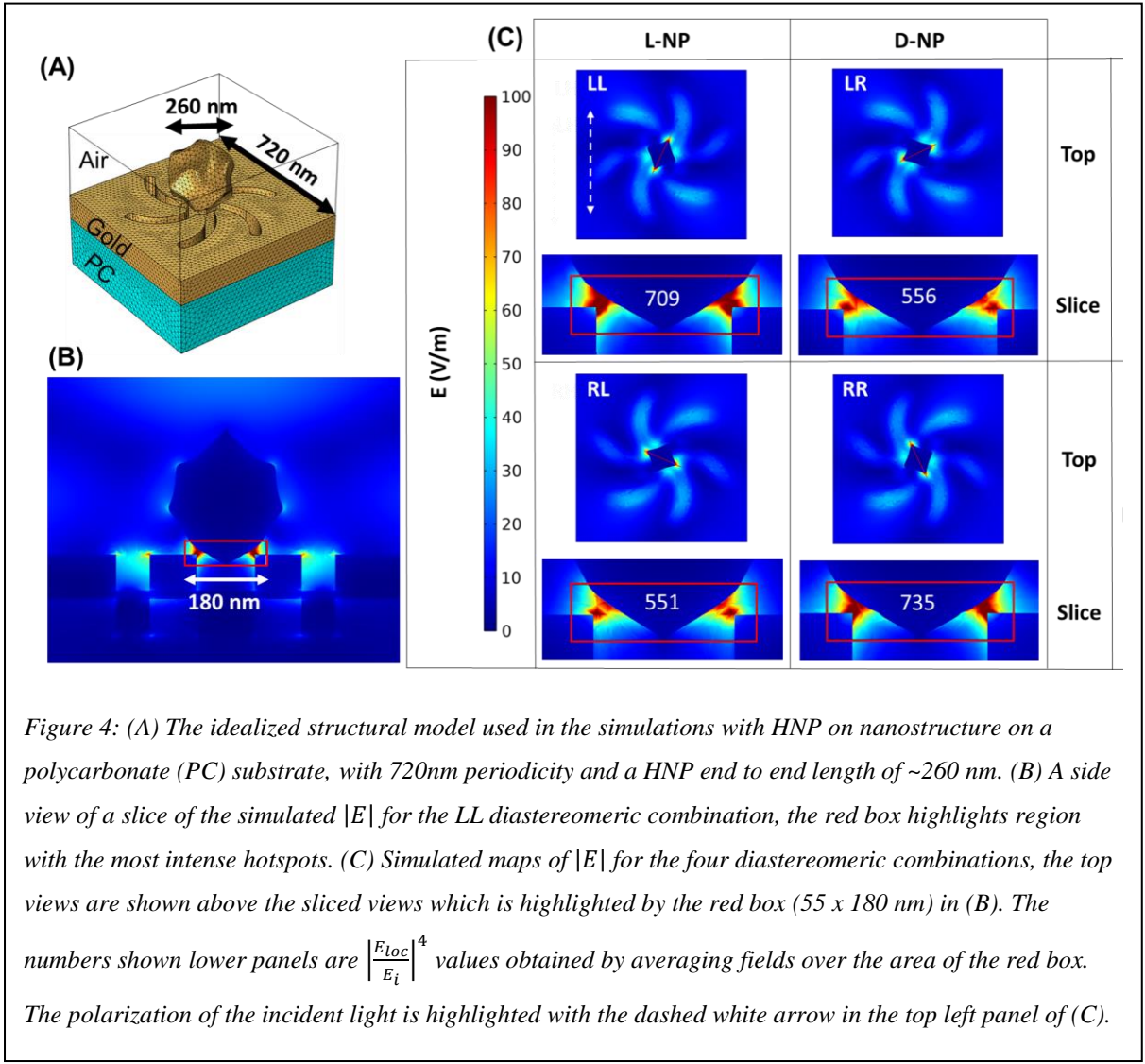

Figure 4: (A) The idealized structural model used in the simulations with HNP on nanostructure on a polycarbonate (PC) substrate, with 720nm periodicity and a HNP end to end length of ~260 nm. (B) A side view of a slice of the simulated |E| for the LL diastereomeric combination, the red box highlights region with the most intense hotspots. (C) Simulated maps of |E| for the four diastereomeric combinations, the top views are shown above the sliced views which is highlighted by the red box (55 x 180 nm) in (B). The numbers shown lower panels are $\left|\frac{E_{loc}}{E_i}\right|^4$ values obtained by averaging fields over the area of the red box. The polarization of the incident light is highlighted with the dashed white arrow in the top left panel of (C).

an argument that can be validated through numerical EM simulations. Given the structural complexity of the deposited layers, simulations of the different symmetry combinations were performed using a single HNP placed in close vicinity to a shuriken nanostructure, Figure 4 (A), as a simplified representation of our experimental samples. When simulating the HNPs on the shuriken nanostructure with a 633 nm excitation, the most intense hotspot, *Figure 4* (B), were found in the narrow gap regions between the shuriken film and the HNP. As would be expected the average $\left|\frac{E_{loc}}{E_i}\right|$ values derived for symmetry related diastereomeric pairs are very similar, *Figure 4*(C). We assign the slight difference in value between mirror symmetry related diastereomers to the computational constraints placed on mesh element size. The simulations clearly show that the most intense hotspots, thus the strongest Raman responses, are produced by diastereomers formed from matched pairs. From the information provided by the simulations we can calculate the ratio of the signal from mis-matched and matched pairs (LR/RR and RL/LL). Assuming the Raman response is dominated by contribution from these hotspots and the intensity of the signal is dependent on $\left|\frac{E_{loc}}{E_i}\right|^4$ values for LR/RR and RL/LL are 0.76 and 0.78. These values are derived by taking an average value in an area defined by the red box in Figure 4(C), which contains the hotspots, but also significant areas within the Au structures where there is no significant field. Consequently, one would expect the actual ratios in the hotspot region to be lower, than have been calculated. Thus, with this caveat, the ratios obtained from the simulations are in reasonable agreement with the experimental observed values of 0.43 and 0.46. So, although based on a highly idealized model the numerical simulations support our hypothesis that diastereomeric pairs, which are not related by mirror symmetry, have different EM near field environments which can be detected using the SERS response from a reporter molecule.

## Conclusion

Before the advent of routine chiroptical characterization, chemists determined the unknown absolute configurations of chiral molecules by combining them with other chiral species of known handedness to produce diastereomers. While the enantiomers of chiral compounds had identical physical and electronic properties, the diastereomers produced from them did not. Thus, allowing physical and spectroscopic methods to discriminate between them, and the absolute configurations to be determined. The work we present is a 21$^{st}$ century EM analogue of this well-established mid-20$^{th}$ century chemical strategy. We have achieved enantiospecific recognition by discriminating between the different near field environments possessed by diastereomers which are not related by mirror symmetry. By using SERS and an appropriate reporter molecule, we have achieved chiral recognition by detecting "cooler" and "hotter" hotspots. The near field sensitivity of SERS makes it a more sensitive probe to local changes in EM environment, than reflectance measured in the far field which monitors a much larger volume of the near field. It should be stressed that the SERS measurements reported here are not intrinsically optically active (i.e., chirally sensitive), and thus contrast with previous work which used metamaterials to amplify the sensitivity of SEROA measurements[8]. The advantage of the approach of leveraging "*meta-diastereomers*" used in this study is that the asymmetric responses are vastly larger than 2 orders of magnitude than are observed in SEROA. Although our study involves a chiral plasmonic HNP the SERS approach we outline would also be valid for detecting local changes in near fields of chiral metafilms induced by chiral molecules. Consequently, the SERS based strategy demonstrated in this work would enable rapid recognition of chiral nanoparticles and compounds using very small sample quantities with applications in drug development and biostructural analysis.

## Methodology

### Fabrication of Chiral Metafilms

The templated substrates were prepared by injection moulding. Clean silicon substrates were coated with ~80 nm of PMMA (Elvacite 2041, Lucite International) and electron beam writing was performed on a Vistec VB6 UHR EWF lithography tool operating at 100 kV. The substrates were then developed using a mixture of methyl isobutyl ketone and isopropyl alcohol to complete the Si master. 300 μm thick nickel shim was formed through electroplating on the Si master. The shim was mounted in a custom-made tool used with an Engel Victory Tech 28 tons injection moulding machine in a fully automatic production mode to manufacture polymer slides using polycarbonate (Makrolon DP2015) as feedstock. The injection moulded substrates have chiral nanostructures imparted in the plastic surface and were subsequently covered by a continuous 100 nm Au film to complete the metafilm sample production.

### Synthesis of Chiral Nanoparticles

Hexadecyltrimethylammonium bromide (CTAB, 99%), L-ascorbic acid (AA, 99%), D-cysteine (99%) L-cysteine (98,5 %), tetrachloroauric(III) trihydrate (HAuCl4·3H2O, 99.9%), biphenyl-4,4′-dithiol (95%) and ultrapure water (LiChrosolv) were purchased from Sigma-Aldrich and used without further purification.

Cubic seeds was synthesized as reported previously.[33,34] Further synthesis was performed according to the published procedure with some modifications.[23] In the next step a growth solution was prepared: 0.8 ml of 100 mM CTAB and 0.2 ml of 10 mM gold chloride trihydrate were added into 3.95 ml of deionized water. Then 0.475 ml of 100 mM ascorbic acid solution was added to the growth solution to reduce $Au^{3+}$ to $Au^{+}$. Further, to prepare chiral nanoparticles, 0.05 ml of cubic seeds were added to the growth solution, and after a 20-minute incubation, 0.005 ml of cysteine were added. The growth solution was left for 2 h at 30 °C, during this time, the solution changed colour from pink to blue. The solution was centrifuged twice (3000 rpm, 180 s) and was dispersed in a 1 mM CTAB solution for further characterization.

**Sample Preparation**

Au-coated shurikens were first rinsed with methanol and plasma cleaned for 5 minutes at 100 W power. Shurikens were then immersed in a 3 mM methanol solution of biphenyl-4,4'-dithiol (95%) (Sigma-Aldrich) for 24 hours to form a self-assembled monolayer (SAM). After rinsing and drying the shurikens with methanol a solution of either L- or D-chiral nanoparticles in DI water was added. The samples were left on a shaker for 24 hours, then washed and dried with DI water.

**Surface characterization techniques.**

SEM images for shurikens were acquired with a SU8240 (Hitachi) using 10 kV acceleration.

**Measurements of optical spectra**.

A custom built polarimeter, equipped with a tungsten halogen light source (Thorlabs), polarizers (Thorlabs), and a 10x objective (Olympus) was used to collect ORD measurements and reflectivity spectra. The samples are positioned and aligned with the aid of a camera (Thorlabs), and the spectrum was measured using a compact spectrometer (Ocean optics USB4000). Reflectivity measurements used plain Au as a background. ORD spectra were obtained using the Stokes' method. Linearly polarized light was positioned incident on the substrate, and the intensities of the reflected light at four polarization angles (0, ±45, and 90°) relative to the incident light were measured to calculate the ORD. The final spectrum is the mean of 8 measurements for a single nanostructure array.

**Raman measurement**

Raman spectra were recorded using a NT-MDT NTEGRA Raman microscope with a 633 nm laser excitation (35mW power) with 10 sec accumulation time. Excitation and collection of Raman scattered light was done using a 20× objective. From each sample, 9 LH and RH shuriken array pairs were measured (18 measurements total measurements each sample). For analysis, the mean and standard error of the SERS intensity values were calculated.

**Numerical Electromagnetic Modelling**

A commercial finite element package (COMSOL V6.0 Multiphysics software, Wave Optics module) was used to simulate the EM fields produced across the sample. Periodic boundary conditions were applied to shuriken boundaries to replicate metamaterial conditions. Perfectly matched layer (PML) conditions were applied to both input and output ports to minimize internal reflections.

We simulated the shuriken metafilms along with a 240 nm diameter chiral L- or D-nanoparticle positioned 250 nm above the shuriken center (Supporting Information Figure S7). Due to computational constraints, the NP could not be positioned 1 nm away from the gold surface (as expected experimentally). Maps showing the distribution of the electric field have been plotted for all the combinations of shuriken and NP handedness. These maps depict where the hot spots (HS) are located between the NP and the metafilm surface. A vertical slice across the HS was obtained, and the average value of the amplitude of the electric field was computed and compared for all the combinations (*Figure 4*).

# Acknowledgement

Martin Kartau and Anastasia Skvortsova would like to acknowledge ScotCHEM and the Scottish Government for funding received under the SFC Saltire Emerging Researcher ScotCHEM European Exchanges Scheme. AK would like to acknowledge support by the UKRI, EPSRC (EP/S001514/1 and EP/S029168/1) and the James Watt Nanofabrication Centre. NG and MK acknowledge support from EPSRC (EP/S012745/1 and EP/S029168/1). MK acknowledges the Leverhulme Trust (RF-2019-023) Oleksiy Lyutakov would like to acknowledge support by the GACR under the project 20-19353S.